\documentclass[11pt]{article}%
\usepackage{makeidx, siunitx, ragged2e}
\usepackage{booktabs,color}
\usepackage{amsfonts,bm}
\usepackage{amsmath}
\usepackage{amssymb}
\usepackage{graphicx}
\usepackage{rotating}
\usepackage{multirow}%
\usepackage{geometry,setspace}
\usepackage{natbib}
\usepackage{multicol}
\usepackage{subfigure}
\usepackage{rotating}
\usepackage{tikz}
\usepackage{pdfpages}
\usepackage{ulem}
\usepackage{booktabs}
\usepackage{lscape}
\usepackage{caption}
\usepackage{lipsum}
\usepackage{longtable}
\newcommand{\E}{\mathbb{E}}

\begin{document}

\title{Global Index on Financial Losses due to Crime in the United States}
\author{
	Thilini Mahanama\thanks{Texas Tech University, Department of Mathematics
		\& Statistics, Lubbock TX 79409-1042, U.S.A., thilini.v.mahanama@ttu.edu (Corresponding
		Author).}
	\and 
	Abootaleb Shirvani\thanks{Texas Tech University, Department of Mathematics
		\& Statistics, Lubbock TX 79409-1042, U.S.A., abootaleb.shirvani@ttu.edu}
	\and 
	Svetlozar Rachev\thanks{Texas Tech University, Department of Mathematics
	\& Statistics, Lubbock TX 79409-1042, U.S.A., Zari.Rachev@ttu.edu}
}
\date{}
\maketitle


\begin{abstract}
Crime can have a volatile impact on investments. 
Despite the potential importance of crime rates in investments, there are no indices dedicated to evaluating the financial impact of crime in the United States. 
As such, this paper presents an index-based insurance portfolio for crime in the United States by utilizing the financial losses reported by the Federal Bureau of Investigation for property crimes and cybercrimes. 
Our research intends to help investors envision risk exposure in our portfolio, gauge investment risk based on their desired risk level, and hedge strategies for potential losses due to economic crashes. 
Underlying the index, we hedge the investments by
issuing marketable European call and put options and 
providing risk budgets (diversifying risk to each type of crime). 
We find that real estate, ransomware, and government impersonation are the main risk contributors. 
We then evaluate the performance of our index to determine its resilience to economic crisis. 
The unemployment rate potentially demonstrates a high systemic risk on the portfolio compared to the economic factors used in this study.
In conclusion, we provide a basis for the securitization of insurance risk from certain crimes that could forewarn investors to transfer their risk to capital market investors.
\end{abstract}




\section{Introduction}\label{sec:Introduction_CI}


The United States spends approximately between 2\% and 6\% of the nation`s gross domestic product on crime victimization \citep{lugo2019estimating}.
The Department of Justice reported that federal, state, and local governments spent more than \$280 billion on criminal justice, including police protection, the court system, and prisons in 2012 \citep{UCR}.
Also, the financial losses due to crime can be viewed as significant dynamic factors that can influence and even shock the financial market in the United States.
In 2019, 6,925,677 property crimes were reported nationwide where the victims suffered losses estimated at \$15.8 billion.

The Uniform Crime Reports published by the Federal Bureau of Investigation outline the financial losses caused by violent crimes (murder, rape, robbery, and aggravated assault) and property crimes (burglary, larceny-theft, motor vehicle theft and arson) \citep{UCR2011}. 
They define the aggregate crime rate as an index (known as UCR Index) for gauging fluctuations in the overall volume and rate of crime \citep{hindelang1974uniform}.
This unweighted index does not utilize the intensities of heterogeneous crimes \citep{kwan2000crime}.  
The Sellin-Wolfgang index addresses this issue by delineating a procedure for adding weights based on severities of crimes 
\citep{blumstein1974seriousness}.
However, it is correlated with the UCR index and provides little additional information.

Over the years, some other types of crimes have contributed immensely to security threats in the United States. 
In particular, cybercrime has grown exponentially with the expansion of internet usage and e-commerce.
In 2019, the Internet Crime Complaint Center of the Federal Bureau of Investigation reported 467,361 complaints - an average of nearly 1,300 every day - with financial losses of more than \$3.5 billion to individual and business victims \citep{ICR}.
Determining the economic impacts caused by these crimes could help policy makers assess the level of systemic risk related to future compensations \citep{lugo2019estimating}.
Thus, this paper proposes an index updated with the impact of all emerging crime types that can be used to more accurately determine the economic impact of crime.


To create this index, we begin by examining the impact of crime on insurance policies in the United States by analyzing the financial losses associated with various types of crimes as reported by the Federal Bureau of Investigation.
The objective of this research is to employ financial methods to construct a reliable and dynamic aggregate index based on economic factors to provide a basis for the securitization of insurance risk from crimes.
Taking all the available data in uniform crime reports and internet crime reports, 
we model the financial losses generated by property crimes and cybercrimes. 
Then, we propose a portfolio based on the economic damages due to crimes
and validate it via value at risk backtesting models.

We hedge the investments underlying the portfolio using two methods
to assess the level of future systemic risk.
First, we issue marketable financial contracts in our portfolio, 
the European call and put options,
to help the investors strategize buying call options and selling put options in our portfolio based on their desired risk level.
Second, we hedge the investment by diversifying risk to each type of crime based on tail risk and center risk measures.
According to the estimated risk budgets, real estate, ransomware, and government impersonation mainly contribute to the risk in our portfolio.
These findings will help investors to envision the amount of risk exposure with financial planning on our portfolio.

We evaluate the performance of our index with respect to economic factors 
to determine the strength of our index by investigating its resilience to the economic crisis.
The unemployment rate potentially demonstrates a high systemic risk on the portfolio compared to the economic factors used in this study.
Therefore, the key findings are intended to provide a basis for the securitization of insurance risk from crime.
This will help insurers gauge investment risk in our portfolio based on their desired risk level and hedge strategies for potential losses due to economic crashes.

Despite the potential significance of crime rates in investments, none of the crime indices researched for this study take economic impacts in to account. 
Our proposed index attempts to address this shortcoming by creating a financial instrument for hedging the intrinsic risk.
This index utilizes the significance of many types of crimes (total of 32) compared to the UCR index. 
Rather than assessing the weights based on severity, we provide risk budgets to identify the potential risk contribution of each crime type.
In conclusion, the main contribution of this research is to provide investors an understanding of how crime can impact insurance risk, which would forewarn and allow them to transfer that risk to capital market investors.

The remainder of this paper's contents are as follows.
First, we model the financial losses due to property crimes and cybercrimes reported by the Federal Bureau of Investigation to propose a portfolio and
backtest it using value at risk models in section \ref{sec:Data_CI}.
In section \ref{sec:OP_CI}, we provide fair values for European option prices and implied volatilities for our portfolio.
Then, we find the risk attributed to each type of crime based on tail risk and center risk measures in section \ref{sec:RB_CI}.
We evaluate the performance of our index with respect to economic factors via stress testing in section \ref{sec:ST_CI}.
Finally, we make concluding remarks in section \ref{sec:DC_CI}.

\section{Financial Losses due to Crime in the United States} \label{sec:Data_CI}


In this section, we propose an index based on the financial losses caused by various types of crimes reported 
between 2001 and 2019 as a proxy to assess the level of future systemic risk caused by crimes.
First, we describe the crime data used in this study in section \ref{sec:DataDescription_CI}.
Then, we model the multivariate time series of financial losses due to crimes in section \ref{sec:TS}.
As a result, we propose a portfolio using the annual cumulative financial losses due to property crimes and cybercrimes.
Finally, we perform backtesting for our index using value at risk models in section \ref{sec:BT_CI}.

\subsection{Crime Data Description}\label{sec:DataDescription_CI}

In this section, we define the types of crimes utilized for constructing our index. 
Using official data published by the Federal Bureau of Investigation (FBI), we considered financial losses caused by crimes committed in the United States between 2001 and 2019.
We use the FBI's Internet Crime Reports 
\cite{ICR} to estimate the financial losses attributed to cybercrimes and
Uniform Crime Reports \cite{UCR} to assess the financial losses caused by property crimes (burglary, larceny-theft, and motor vehicle theft).
Using the information collected from these two reports, we calculate the cumulative financial losses reported for the following 32 types of crimes \cite{UCR2011,ICR}:

\begin{itemize}
	
	\item \textbf{Advanced Fee:}
	An individual pays money to someone in anticipation of receiving something of greater value in return, but instead receives significantly less than expected or nothing.
	\item \textbf{BEA/EAC (Business Email Compromise/Email Account Compromise)}: BEC is a scam targeting businesses working with foreign suppliers and/or businesses regularly performing wire transfer payments. EAC is a similar scam that targets individuals. These sophisticated scams are carried out by fraudsters compromising email accounts through social engineering or computer intrusion techniques to conduct unauthorized transfer of funds.
	\item \textbf{Burglary:}
	The unlawful entry of a structure to commit a felony or a theft. Attempted forcible entry is included.
	\item \textbf{Charity:}
	Perpetrators set up false charities, usually following natural disasters, and profit from individuals who believe they are making donations to legitimate charitable organizations.
	\item \textbf{Check Fraud:}
	A category of criminal acts that involve making the unlawful use of cheques in order to illegally acquire or borrow funds that do not exist within the account balance or account-holder's legal ownership.
	\item \textbf{Civil Matter:}
	Civil lawsuits are any disputes formally submitted to a court that is not criminal.
	\item \textbf{Confidence Fraud/Romance:}
	A perpetrator deceives a victim into believing the perpetrator
	and the victim have a trust relationship, whether family, friendly or romantic. As a result of that belief, the victim is persuaded to send money, personal and financial information, or items of value to the perpetrator or to launder money on behalf of the perpetrator. Some variations of this scheme are romance/dating scams or the grandparent scam.
	\item \textbf{Corporate Data Breach:}
	A leak or spill of business data that is released from a secure location to an untrusted environment. It may also refer to a data breach within a corporation or business where sensitive, protected, or confidential data is copied, transmitted, viewed, stolen, or used by an individual unauthorized to do so.
	\item \textbf{Credit Card Fraud:}
	Credit card fraud is a wide-ranging term for fraud committed using a credit card or any similar payment mechanism as a fraudulent source of funds in a transaction.
	\item \textbf{Crimes Against Children:}
	Anything related to the exploitation of children, including child abuse. 
	\item \textbf{Denial of Service:}
	A Denial of Service (DoS) attack floods a network/system or a Telephony Denial of Service (TDoS) floods a service with multiple requests, slowing down or interrupting service.
	\item \textbf{Employment:}
	Individuals believe they are legitimately employed, and lose money or launders money/items during the course of their employment.
	\item \textbf{Extortion:}
	Unlawful extraction of money or property through intimidation or undue exercise of authority. It may include threats of physical harm, criminal prosecution, or public exposure.
	\item \textbf{Gambling:}
	Online gambling, also known as Internet gambling and iGambling, is a general term for gambling using the Internet.
	\item \textbf{Government Impersonation:}
	A government official is impersonated in an attempt to collect money.
	\item \textbf{Harassment/Threats of Violence:}
	Harassment occurs when a perpetrator uses false accusations or statements of fact to intimidate a victim. Threats of Violence refers to an expression of an intention to inflict pain, injury, or punishment, which does not refer to the requirement of payment.
	\item \textbf{Identity Theft:}
	Identify theft involves a perpetrator stealing another person’s personal identifying information, such as name or Social Security number, without permission to commit fraud.
	\item \textbf{Investment:}
	A deceptive practice that induces investors to make purchases on the basis of false information. These scams usually offer the victims large returns with minimal risk.
	Variations of this scam include retirement schemes, Ponzi schemes, and pyramid schemes.
	\item \textbf{IPR Copyright:}
	The theft and illegal use of others’ ideas, inventions, and creative expressions, to include everything from trade secrets and proprietary products to parts, movies, music, and software.
	\item \textbf{Larceny Theft:}
	The unlawful taking, carrying, leading, or riding away of property (except motor vehicle theft) from the possession or constructive possession of another.  
	\item \textbf{Lottery/Sweepstakes:}
	Individuals are contacted about winning a lottery or sweepstakes they never entered, or to collect on an inheritance from an unknown relative and are asked to pay a tax or fee in order to receive their award.
	\item \textbf{Misrepresentation:}
	Merchandise or services were purchased or contracted by individuals online for which the purchasers provided payment. The goods or services received were of measurably lesser quality or quantity than was described by the seller.
	\item \textbf{Motor Vehicle Theft:}
	The theft or attempted theft of a motor vehicle.  A motor vehicle is self-propelled and runs on land surface and not on rails.  Motorboats, construction equipment, airplanes, and farming equipment are specifically excluded from this category.
	\item \textbf{Non-Payment/Non-Delivery:}
	In non-payment situations, goods and services are shipped, but payment is never rendered. In non-delivery situations, payment is sent, but goods and services are never received.
	\item \textbf{Overpayment:}
	An individual is sent a payment/commission and is instructed to keep a portion of the payment and send the remainder to another individual or business.
	\item \textbf{Personal Data Breach:}
	A leak or spill of personal data that is released from a secure location to an untrusted environment. It may also refer to a security incident in which an individual's sensitive, protected, or confidential data is copied, transmitted, viewed, stolen, or used by an unauthorized individual.
	\item \textbf{Phishing/Vishing/Smishing/Pharming:}
	Unsolicited email, text messages, and telephone calls purportedly from a legitimate company requesting personal, financial, and/or login credentials.
	\item \textbf{Ransomware:}
	A type of malicious software designed to block access to a computer system until money is paid.
	\item \textbf{Real Estate/Rental:}
	Fraud involving real estate, rental, or timeshare property.
	\item \textbf{Robbery:}
	The taking or attempting to take anything of value from the care, custody, or control of a person or persons by force or threat of force or violence and/or by putting the victim in fear.
	\item \textbf{Social Media:}
	A complaint alleging the use of social networking or social media (Facebook, Twitter, Instagram, chat rooms, etc.) as a vector for fraud. Social Media does not include dating sites.
	\item \textbf{Terrorism:}
	Violent acts intended to create fear that are perpetrated for a religious, political,
	or ideological goal and deliberately target or disregard the safety of non-combatants.
	
\end{itemize}

Whenever necessary, we use multiple imputations with the principal component analysis model to compute missing data \cite{josse2011multiple}. 
Moreover, we adjust the financial losses for U.S. dollars in 2020 using the CPI Inflation Calculator 
available in the U.S. Bureau of Labor Statistics.
Then, we model the time series of financial losses due to these crime types in section \ref{sec:TS}.\\

\subsection{Modeling the Multivariate Time Series of Financial Losses due to Crimes}\label{sec:TS}

In this section, we model the financial losses due to the 32 types of crimes described in section \ref{sec:Data_CI}.
In each type of crime, we transform the series of financial losses to a stationary time series 
by taking the log returns:

\begin{equation} \label{Eq:logReturns}
R_t^{(i)}=\log L_t^{(i)}-\log L_{t-1}^{(i)}; \;\;\;\; i=1,..32, \;\;\; t=0,..,T
\end{equation}

\noindent where $L_t^{(i)}$ denotes the financial loss due to $i$\textsuperscript{th} crime type at time $t$.

Then, we fit the Normal Inverse Gaussian (NIG) distribution to each log return $r_t^{(i)}$ series
and estimate parameters using the maximum likelihood method.
As a result, we have NIG L\'{e}vy processes for each type of crime.
Moreover, since NIG has an exponential form at the moment generating function, we use these dynamic returns for option pricing in section \ref{sec:OP_CI}.
Then, for each NIG L\'{e}vy process, we generate 10,000 scenarios to obtain independent and identically distributed data for returns.






\subsection{Backtesting the Portfolio} \label{sec:BT_CI}

In this study, we propose a portfolio based on the annual cumulative financial losses due to all types of crimes described in section \ref{sec:DataDescription_CI}.
Then, we convert this portfolio to a stationary time series by taking their log returns.
We denote $r_{t}$ as the log return of the index at time $t$ where $\mu_{t}$ is drift and  $\sigma_{t}$ is volatility:

\begin{equation} \label{Eq:mu&a}
r_{t}=\mu_{t}+a_{t}, \;\;\; t=0, \ldots, T.
\end{equation}

\noindent Then, we model the log-returns using the ARMA(1,1)-GARCH(1,1) filter to eliminate the serial dependence.
In particular, we use ARMA(1,1) \cite{whittle1953analysis} to model the drift ($\mu_{t}$)

\begin{equation} \label{Eq:ARCH}
\mu_{t}=\phi_{0}+\phi_{1} r_{t-1}+\theta_{1} a_{t-1}
\end{equation}

\noindent and GARCH(1,1) \cite{bollerslev1986generalized} to model the volatility ($\sigma_{t}$)

\begin{equation} \label{Eq:GARCH}
\begin{split}
\sigma_{t} &=\frac{a_{t}}{\epsilon_{t}} \\
\sigma_{t}^{2} &=\alpha_{0}+\alpha_{1}a_{t-1}^{2}+\beta_{1}\sigma_{t-1}^{2}
\end{split}
\end{equation}	

\noindent where $\phi_{0}$ and $\alpha_{0}$ are constants and $\phi_{1}, \theta_{1}, \alpha_{1}$ and $\beta_{1}$ are parameters to be estimated. Moreover, the sample innovations, $\epsilon_{t}$, follows an arbitrary distribution with zero mean and unit variance.\\

In particular, we assume Student's t and NIG for the distributions of sample innovations, $\epsilon_{t}$.
Then, we examine the performance of these two filters (ARMA(1,1)-GARCH(1,1) with Student's t innovations and ARMA(1,1)-GARCH(1,1) with NIG innovations) via backtesting.
Furthermore, we utilize the better model obtained in this section to implement option pricing to our portfolio in section \ref{sec:OP_CI}.

We backtest the models using Value at Risk (VaR) measures \cite{jorion2007value}.
In VaR backtesting \cite{nieppola2009backtesting}, we compare the actual returns with the corresponding VaR models.
The level of difference between them helps to identify whether the VaR model is underestimating or overestimating the risk. 
Moreover, if the total failures are less than expected, then the model is considered to overestimate the VaR, and if the actual failures are greater than expected, the model underestimates VaR.

To perform backtesting, we use the residuals of filters between 2002 and 2015 to train the model.
Then, the test window starts in 2016 and runs through the end of the sample (2019).
We perform the backtesting for the out-of-sample data at the quantile levels of 0.01, 0.05, 0.25, 0.50, 0.75, 0.95, and 0.99.
For the $\alpha$ quantile level, we define VaR as follows:

\begin{equation} \label{Eq:VaR}
\text{VaR}_\alpha(x) = -\inf \{x \; | \; F(x)>\alpha, \;  x \in \mathbb{R}\}
\end{equation}

\noindent where $F(x)$ is the cumulative density function of the returns.

\begin{table}[]
	\caption{VaR Backtesting Results for ARMA(1,1)-GARCH(1,1) with Student's t and NIG innovations.}
	\centering
	\begin{tabular}{@{}lccccc@{}}
		\toprule
		\multicolumn{1}{l}{\multirow{2}{*}{Innovation}} & \multirow{2}{*}{VaR Level} & \multicolumn{3}{c}{Test Results} \\ \cmidrule(l){3-6} 
		\multicolumn{1}{c}{} & & Traffic Light & Binomial & PoF & CCI\\ \cmidrule(r){1-6}
		\multirow{3}{*}{Student's t} 
		& 0.01 & green & reject & reject & accept\\
		& 0.05 & green & accept & accept& accept\\
		& 0.25 & green & reject & reject& accept\\
		& 0.50 & green & accept & accept& accept\\
		& 0.75 & green & accept & accept& accept\\
		& 0.95 & green & accept & accept& accept\\
		& 0.99 & yellow & accept & accept& accept\\
		\hline
		\multirow{3}{*}{NIG}
		& 0.01 & green & reject & reject& accept\\
		& 0.05 & green & reject & reject& accept\\
		& 0.25 & green & reject & reject& accept\\
		& 0.50 & green & accept & accept& accept\\
		& 0.75 & yellow & reject & reject& accept\\
		& 0.95 & red & reject & reject& accept\\
		& 0.99 & red & reject & reject& accept\\
		\hline
	\end{tabular}
	\label{tab:VaR-Backtesting}
\end{table}

Table \ref{tab:VaR-Backtesting} provides the results for VaR backtesting on ARMA(1,1)-GARCH(1,1) with Student's t and NIG filters.
First, we perform the Conditional Coverage Independence (CCI) to test for independence \cite{braione2016forecasting}.
According to Table \ref{tab:VaR-Backtesting}, both filters show independence on consecutive returns.

Then, we perform traffic light, binomial test, and proportion of failures (PoF) tests as frequency tests.
ARMA(1,1)-GARCH(1,1) with Student's t model is generally acceptable in the frequency tests at most of the levels.
However, ARMA(1,1)-GARCH(1,1) with NIG model fails at all the levels except the 0.5 level.

In conclusion, ARMA(1,1)-GARCH(1,1) with Student's t innovations outperforms the ARMA(1,1)-GARCH(1,1) with NIG model in backtesting.
Hence, we utilize ARMA(1,1)-GARCH(1,1) with Student's t to implement option pricing in section \ref{sec:OP_CI}.

\section{Option Prices for the Crime Portfolio} \label{sec:OP_CI}

An option is a contract between two parties that gives one party the right, but not the obligation, to buy or sell the underlying asset at a prespecified price within a specific time.  
We provide fair prices of call and put options in our portfolio in section \ref{sec:OP_Results} 
based on the pricing model defined in section \ref{sec:OP_Model}.
Then, we investigate the implied volatilities of our index using the Black-Scholes and Merton model. 
Ultimately, the findings of this section are intended to help investors strategize buying call options and selling put options of our portfolio based on their desired risk level and predicted volatilities.

\subsection{Defining a Model for Pricing Options} \label{sec:OP_Model}

The theoretical value of an option estimates its fair value based on strike price and time to maturity\footnote{See \cite{black2019pricing, madan1989multinomial, hurst1999option, carr2004time, bell2006option, klingler2013option, rachev2017option} and \cite{shirvani2020option}}. 
In pricing options, the conventional Black-Scholes Model\footnote{See \cite{Black_1973} and \cite{Merton_1973}} assumes the price of a financial asset follows a stochastic process based on a Brownian motion with a normal distribution assumption.
Since the asset returns are heavy-tailed in practice, the extreme variations in prices cannot be well captured using a normal distribution.
With non-normality assumption, we implement a L\'{e}vy processes for asset returns.
This provides better estimates for prices since the non-marginal variations are more likely to happen as a consequence of fat-tailed distribution-based processes\footnote{See \cite{mandelbrot1967distribution, clark1973subordinated, ken1999levy, hurst1997subordinated}, and \cite{shirvani2020multiple}}.

Among L\'{e}vy processes, the NIG process \cite{Barndorff_1997} is widely used for pricing options 
as it allows for wider modeling of skewness and kurtosis than the Brownian motion does. 
Thus, it enables us to estimate consistent option prices with different strikes and maturities using a single set of parameters. 
We use the NIG process to price options for the crime portfolio based on the estimated parameters for the returns obtained using the maximum likelihood estimation given in Table \ref{tab:NIG_par}.

\begin{table}[h!]
	\caption{The estimated parameters of the fitted the NIG process to the Crime portfolio log-returns}
	\centering
	\begin{tabular}{@{}lcccc@{}}
		\toprule
		Parameters & $\mu$      & $\alpha$  & $\beta$   & $\delta$  \\ \midrule
		Estimates  & -0.0014 & 0.4826 & 0.0006 & 0.6553 \\ \bottomrule
	\end{tabular}
	\label{tab:NIG_par}
\end{table}

The NIG process is a Brownian motion where the time change process follows an Inverse Gaussian (IG) distribution,
i.e., the NIG process is a Brownian motion subordinated to an IG process. 
We define the NIG process ($X_t$) as a Brownian motion ($B_t$) with drift  
($\mu$) and volatility ($\sigma$)  as follows

\begin{equation}
X_t=\mu t+\sigma B_t, \ \ \ t\ge 0,\ \ \ \mu \in \mathbb{R}, \ \ \ \sigma > 0.
\end{equation}

\noindent At $t=1$, we denote the process as $X_1\sim NIG\left( \mu,\alpha,\beta,\delta\right) $ with parameters $\mu, \alpha,\beta,\delta \in \mathbb{R}$ such that $\alpha^2>\beta^2$ and the density given by
\begin{equation}
\label{NIG_dis}
f_{X_1} \left( x\right) = \frac{\alpha \delta K_1 \left( \alpha \sqrt{ \delta^2 + \left( x-\mu \right) ^2 }\right) } { \pi \sqrt{\delta^2+\left( x-\mu\right) ^2 }} \exp {\left(  \delta \sqrt{ \alpha^2-\beta^2 }+\beta \left( x-\mu \right)\right) } , \,\,x \in \mathbb{R}.
\end{equation}
\noindent The characteristic function of the NIG process is derived using $\varphi_{X_t} \left( t\right) =E\left( e^{itX_t}\right), \, t\in \mathbb{R} $ and is given by
\begin{equation}
\label{NIG_Char}
\varphi_{X_t}\left( t\right)=\exp\left( {i\mu t +\delta\left( \sqrt{\alpha^2-\beta^2}-\sqrt{\alpha^2 - \left( \beta+it\right)^2 }\right) }\right). 
\end{equation}

For pricing financial derivatives, we search for risk-neutral probability ($\mathbb{Q}$) known as Equivalent Martingale Measure (EMM).
The current value of an asset is exactly equal to the discounted expectation of the asset at the risk-free rate under $\mathbb{Q}$. 
In particular, we use Mean-Correcting Martingale Measure (MCMM) for $\mathbb{Q}$ as it is sufficiently flexible for calibrating market data. 
In MCMM, the price dynamics of the price process ($S_t$) on $\mathbb{Q}$ is given by
\begin{equation}
\label{Eq3}
S^{(\mathbb{Q})}_t=\frac{e^{-rt}S_t}{M_{X_t} (1)}, \ \ \ t \geq 0
\end{equation}
\noindent where $M_{X_t}()$ is the moment generating function of $X_t$ and $r$ is the risk free rate. 
We model the risk-neutral log stock-price process for a given option pricing formula and our market model on $\mathbb{Q}$ as follows
\begin{equation}
\label{Eq7}
\begin{split}
{\varphi }_{\ln S^{ (\mathbb{Q} )}_t} (v )=S^{iv}_0 \ {\mathrm{exp}  \{ [iv (r-\ln {\varphi }_{ X_1}(-i))+\ln {\varphi }_{ X_1}(v) ]t\}}
\end{split}
\end{equation}	
where ${\varphi }_{ X_t} (v)=\mathbb{E}(e^{ivX_t})$ is the characteristic function of $X_t$ in Eq.~\eqref{NIG_Char}.

We obtain the EMM using MCMM as the pricing formula is arbitrage-free for the European call option pricing formula under $\mathbb{Q}$.
First, we estimate all the parameters involved in the process and add the drift term, $X^{new}_t = X^{old}_t + mt$, in such a way that the discounted stock-price process becomes a martingale.

We define the price of a European call contract ($\mathcal{C}$) with underlying risky assets at $t=0$ as
\begin{equation}
\label{Eq5}
C (S_0,r,K,T )=e^{-rT} \ {\mathbb{E}}^{\mathbb{Q}} \ {\mathrm{max}  (S^{ (\mathbb{Q} )}_T-K,0 )}, \ \ \ K,T>0, 
\end{equation}
for the given price process $(S_t)$, time to maturity ($T$), and strike price ($K$).  

When the characteristic function of the risk-neutral log stock-price process is known, Carr and Madan's study \cite{Carr_1999} derives the pricing method for the European option valuation using the fast Fourier transform.
Following that, we price the European options contract with an underlying risky asset using the characteristics function, Eq (\ref{NIG_Char}), and fast Fourier transform to convert the generalized Fourier transform of the call price.
For any positive constant $a$ such that ${\mathbb{E}}^{\mathbb{Q}}{ (S^{ (\mathbb{Q} )}_T )}^{a}<\infty $ exists, we define the call price as

\begin{equation}
\label{Eq6}
C (S_0,r,K,T )=\frac{e^{-rT-ak}}{\pi }\int^{\infty }_0{e^{-ivk}}\frac{{\varphi }_{\ln S^{ (\mathbb{Q} )}_T} (v-i(a+1) )}{a(a +1) -v^2+2i (a +1)v} \ dv, \ \ \ a>0,
\end{equation}
where $k=\ln K$ and ${\varphi }_{\ln S^{ (\mathbb{Q} )}_t} (v )$ is the characteristic function  of the log-price process under $\mathbb{Q}$.  
By utilizing call option prices in the put–call parity formula, we calculate the price of a put option in the crime portfolio.

\subsection{Issuing the European Option Prices for the Crime Portfolio} \label{sec:OP_Results}

In this section, we calculate the European call and put option prices for our portfolio using the pricing model, Eq (\ref{Eq6}), introduced in section \ref{sec:OP_Model}. 
We provide European option prices by fixing $S_0$ to 100 in Eq (\ref{Eq6}), i.e., the price of the crime portfolio at time zero is 100 units,
and the time to maturity is in days.
Later, we provide the implied volatilities of the portfolio based on the volatilities of call and put option prices.

\begin{figure}
	\centering
	\includegraphics[width=0.7\textwidth]{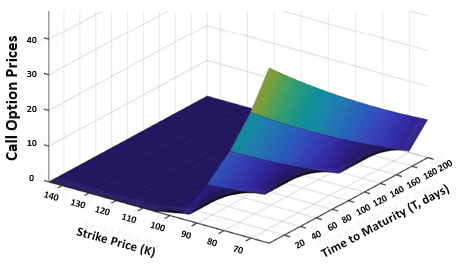}
	\caption{Call option prices against time to maturity ($T$, in days) and strike price ($K$, based on $S_0 = 100$).}
	\label{call_price}
\end{figure}

\begin{figure}
	\centering
	\includegraphics[width=0.7\textwidth]{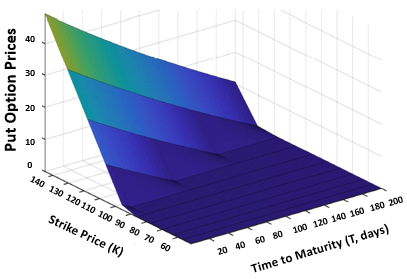}
	\caption{Put option prices against time to maturity ($T$, in days) and strike price ($K$, based on $S_0 = 100$).}
	\label{put_price}
\end{figure}

First, we demonstrate the relationship between call option prices, strike price $(K)$, and time to maturity $(T)$ in Figure \ref{call_price}. 
These calculated prices help the investors to strategize buying the stocks in our portfolio at a predefined price $(K)$ within a specific time frame $(T)$.
Second, we show put option prices in Figure \ref{put_price} 
to provide selling prices of the shares in our index.
The prices of our options validate the fact that option prices decrease as the time to maturity increases for a given strike price.

\begin{figure}
	\centering
	\includegraphics[width=0.7\textwidth]{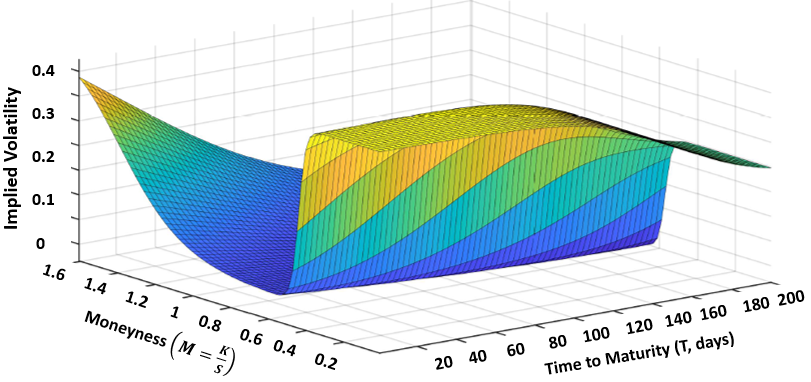}
	\caption{Implied volatility surface against time to maturity ($T$, in days) and moneyness ($M = K/S$, the ratio of strike price, $K$, and stock price, $S$).}
	\label{Implied_put}
\end{figure}

Third, we determine the implied volatilities of our portfolio using the Black-Scholes and Merton model. 
This provides the expected volatility of our portfolio over the life of the option $(T)$.
Figure \ref{Implied_put} is the implied volatility surface with respect to moneyness ($M$)  and time to maturity ($T$).
In particular, we calculate moneyness as the ratio of the strike price ($K$) and stock price ($S$), i.e., $M = K/S$.
Then, the volatilities for call and put option prices are shown in the regions of the surface with $M<1$ and $M>1$, respectively.
The volatility surface demonstrates a volatility smile which is usually seen in the stock market.

Figure \ref{Implied_put} illustrates that implied volatility increases when the moneyness is further out of the money or in the money, compared to at the money $(M=1)$. 
In this case, volatility seems to be low, with a range of 0.8 and 1.2 in moneyness, compared to the other regions in the implied volatility surface,
i.e., the options with higher premiums result in high implied volatilities. 
These findings help investors strategize buying call options and selling put options in our portfolio based on their desired risk level and predicted volatilities.

\section{Risk Budgets for the Crime Portfolio} \label{sec:RB_CI}

The investors intend to optimize portfolio performance while maintaining their desired risk tolerance level \cite{mahanama2020natural}.
In this section, we provide a rationale for investors to determine the degree of variability in our portfolio.
Therefore, we provide the risk contribution related to each type of crime in section \ref{sec:CalculateRisk}
using the risk measures defined in section \ref{sec:RiskMeasures}.
As a result, we present the main risk contributors and risk diversifiers in our portfolio.
Ultimately, these risk budgets (the estimated risk allocations) will potentially help investors with their financial planning (maximizing the returns).

\subsection{Defining Tail and Center Risk Measures} \label{sec:RiskMeasures}

This section defines the risk measures that we use for assessing risk allocations in section \ref{sec:CalculateRisk}.
We determine the tail risk contributors and center risk contributors using expected tail loss and volatility risk measures, respectively \cite{Chow:2001,Boudt:2013}.

We use Conditional Value at Risk (CoVaR) \cite{adrian2011covar, girardi2013systemic} for finding the tail risk contributors in our portfolio at levels of 95\% and 99\%.
We define the tail risk contribution of the $i$\textsuperscript{th} asset at $\alpha$ level as follows:

\begin{equation} \label{Eq:TR}
TR_i(\alpha) = \text{CoVaR}_\alpha^{(i)} =\frac{1}{\alpha} \int_{0}^{\alpha} VaR_{\gamma}^{(i)}(x) \; d \gamma.
\end{equation}

In order to find center risk contributors, we measure the volatility of asset prices using standard deviation. 
Since we utilize an equally-weighted portfolio, we denote the weight vector for 32 types of crimes as $\textbf{w} = (w_1,\cdots,w_{32})$ where $w_i = \frac{1}{32}$. 
We define the volatility risk measure, $R(w)$, using the covariance matrix, $\Sigma$, of asset returns (\ref{Eq:logReturns}) \cite{hu2019modelling}:

\begin{equation} \label{Eq:R(w)}
R(\textbf{w}) = \sqrt{\textbf{w}'\Sigma \textbf{w}}.
\end{equation}

\noindent Then, the center risk contribution of the $i$\textsuperscript{th} asset is given by
\begin{equation} \label{Eq:CR}
CR_i(\textbf{w}) = w_i \frac{\partial R(\textbf{w})}{\partial w_i}.
\end{equation}

Having outlined the risk measures, section \ref{sec:CalculateRisk} utilizes tail and center risk contributions to find the risk allocation for each type of crime defined in section \ref{sec:DataDescription_CI}.

\subsection{Determining the Risk Budgets for the Crime Portfolio} \label{sec:CalculateRisk}

This section assesses the risk attributed to each type of crime using the risk measures defined in section \ref{sec:RiskMeasures}.
First, we calculate the tail risk allocations ($TR$) using Eq (\ref{Eq:TR}) to find the main tail risk contributors in our portfolio.
Then, we investigate the main center risk contributors in our index using center risk allocations ($CR$) computed using Eq (\ref{Eq:CR}).
Finally, taking the main tail risk and center risk contributors into account, we find the main risk contributors of our portfolio.

\begin{table}[]
	\centering
	\caption{The percentages of center risk (CR) and tail risk (TR) (at levels of 95\% and 99\%) budgets for the portfolio on crimes.}
	\begin{tabular}{l r r r} 
		\toprule
		Crime Type 					& $\%TR(95)$ & $\%TR(99)$ & $\%CR$\\ [0.5ex] 
		\hline
		Real Estate					&	13.62	&	12.93	&	11.29	\\
		Ransomware					&	10.35	&	11.40	&	8.48	\\
		Government Impersonation	&	9.48	&	8.80	&	8.25	\\
		Identity Theft				&	7.96	&	10.23	&	6.74	\\
		Extortion					&	7.72	&	6.90	&	7.19	\\
		Lottery						&	7.09	&	7.37	&	6.26	\\
		Confidence Fraud			&	5.56	&	6.64	&	5.48	\\
		Investment					&	5.31	&	7.11	&	4.90	\\
		Crimes Against Children		&	5.24	&	3.95	&	5.55	\\
		Personal Data Breach		&	3.69	&	4.29	&	3.24	\\
		Credit Card Fraud			&	3.58	&	3.16	&	3.96	\\
		BEC/EAC						&	3.27	&	3.29	&	3.00	\\
		Non-Payment					&	2.56	&	4.38	&	2.10	\\
		IPR Copyright				&	2.02	&	2.04	&	1.95	\\
		Gambling					&	1.97	&	2.55	&	1.40	\\
		Robbery						&	1.80	&	0.67	&	6.77	\\
		Phishing					&	1.48	&	0.97	&	2.01	\\
		Civil Matter				&	1.30	&	-0.57	&	2.89	\\
		Denial Of Service			&	1.02	&	-0.23	&	2.74	\\
		Motor Vehicle Theft			&	1.01	&	1.30	&	0.73	\\
		Check Fraud					&	0.98	&	2.19	&	-0.51	\\
		Advanced Fee				&	0.75	&	0.43	&	1.04	\\
		Harassment					&	0.74	&	0.10	&	1.14	\\
		Corporate Data Breach		&	0.70	&	0.03	&	0.90	\\
		Larceny Theft				&	0.50	&	0.56	&	0.39	\\
		Terrorism					&	0.36	&	-0.20	&	2.06	\\
		Burglary					&	0.30	&	0.21	&	0.44	\\
		Employment					&	0.22	&	0.09	&	0.29	\\
		Charity						&	0.17	&	0.18	&	0.54	\\
		Overpayment					&	-0.04	&	-0.23	&	0.10	\\
		Social Media				&	-0.16	&	-0.17	&	-0.31	\\
		Misrepresentation			&	-0.55	&	-0.41	&	-1.00	\\
		\bottomrule
		\label{Tab:RB_CI}
	\end{tabular}
\end{table}

In Table \ref{Tab:RB_CI}, we provide the center risk ($CR$) and tail risk ($TR$) allocations for each type of crime.
We find the risk diversifiers in the portfolio using the negative risk allocations shown in Table \ref{Tab:RB_CI}.
With significantly low center and tail risk diversifications, Misrepresentation and Social Media seems to be the potential main risk diversifiers in our portfolio.

We consider the positive values outlined in Table \ref{Tab:RB_CI} to identify the main risk contributors in our portfolio.
We find the main tail risk contributors using the positive tail risk estimates at levels of 95\% and 99\%.
At the 95\% level, $TR(95\%)$, Real Estate, Ransomware, and Government Impersonation provide a relatively higher tail risk than the other factors.
However, Real Estate, Ransomware, and Identity Theft seem to be the main tail risk contributors at the 99\% level, $TR(99\%)$.

We determine the main center risk contributors in our portfolio using the positive center risk estimates, $CR$, illustrated in Table \ref{Tab:RB_CI}.
Since Real Estate, Ransomware, and Government Impersonation demonstrate high volatility compared to the other types of crimes, they seem to be the main center risk contributors in our portfolio.

As Real Estate and Ransomware are both main tail risk and center risk contributors, they are the potential main risk contributors in our portfolio.
These estimated risk budgets and the main risk contributors will help investors to envision the amount of risk exposure with financial planning on our portfolio.

\section{Performance of the Crime Portfolio for Economic Crisis} \label{sec:ST_CI}

We evaluate the performance of our portfolio using economic factors related to low income
as they are known to be major root causes of crime. 
To investigate the robustness of the crime portfolio for inevitable economic crashes, we perform stress testing in section \ref{sec:CalculateSystemicRisk} based on the systemic risk measures defined in section \ref{sec:SystemicRisk}
The findings of this section are intended to help to determine portfolio risks and serves as a tool for hedging strategies required to mitigate inevitable economic crashes.

\subsection{Defining Systemic Risk Measures}\label{sec:SystemicRisk}

In this section, we define the systemic risk measures used for stress testing the portfolio on crimes.
We define three derived risk measures based on VaR (\ref{Eq:VaR}) denoting $Y$ as the portfolio and $X$ as a stress factor \cite{trindade2020socioeconomic}.

CoVaR is a coherent measure of tail risk in an investment portfolio.
In our study, we use a variant of CoVaR defined in terms of copulas \cite{Mainik14}. 
Using the condition $X\leq \text{VaR}_{\alpha}(X)$ rather than the traditional CoVaR condition, $X= \text{VaR}_{\alpha}(X)$, improves the response to dependence between $X$ and $Y$.
We define CoVaR at level $\alpha$ as

\begin{equation}\label{Eq:CoVaR}
\text{CoVaR}_\alpha = \text{VaR}_{\alpha}\left(Y \; | \; X\leq\text{VaR}_{\alpha}(X)\right).
\end{equation}

The CoVaR for the closely associated expected shortfall is defined as the tail mean beyond VaR \cite{Mainik14}.
Furthermore, we use an extension of CoVaR denoted as Conditional Expected Shortfall (CoES).
Then, we define CoES at level $\alpha$ as follows:

\begin{equation}\label{Eq:CoES}
\text{CoES}_\alpha = \E\left(Y \; | \; Y \;\leq\text{CoVaR}_\alpha, \; X\leq\text{VaR}_{\alpha}(X)\right).
\end{equation}

Conditional Expected Tail Loss (CoETL) \cite{ZariCOETL} is the average of the portfolio losses when all the assets are in distress.
CoETL is an appropriate risk measure to quantify the portfolio downside risk in the presence of systemic risk.
We denote CoETL at level $\alpha$ as

\begin{equation}\label{Eq:CoETL}
\text{CoETL}_\alpha = \E\left(Y \; | \; Y\leq\text{VaR}_{\alpha}(Y), \; X\leq\text{VaR}_{\alpha}(X)\right).
\end{equation}

\noindent We quantify the market risk of our portfolio on crime using these systemic risk measures in section \ref{sec:CalculateSystemicRisk}.

\subsection{Evaluating the Performance of the Crime Portfolio for Economic Factors} \label{sec:CalculateSystemicRisk}

In this section, we evaluate how well our portfolio would perform with economic factors related to low income.
In particular, we test the impact of the Unemployment Rate, Poverty Rate, and Median Household Income on our index.
We quantify the potential impact of these economic factors on our index using the systemic risk measures defined in section \ref{sec:SystemicRisk}.
Since the stress testing results indicate the investment risk in our portfolio,
the investors can utilize the outcomes to hedge strategies for forthcoming economic crashes.

Based on backtesing results in section \ref{sec:BT_CI}, we use the ARMA(1,1)-GARCH(1,1) model with Student-t innovations for the log returns of our portfolio.
Also, we apply this filter to log returns of the economic factors to eliminate inherent linear and nonlinear dependencies.
Then, we fit bivariate NIG models to the joint distributions of independent and identically distributed standardized residuals of each economic factor and our portfolio on crime.
Using these bivariate models, we generate 10,000 simulations for each joint density to perform a scenario analysis.
In Table \ref{Tab:Correlation}, we provide the empirical correlation coefficients of each simulated joint density with the corresponding economic factor.
This table demonstrates weak correlations between the economic factors and the portfolio.

\begin{table}
	\caption{The empirical correlation coefficients of the joint densities of each economic factors and the crime portfolio}
	\centering
	\begin{tabular}{l r} 
		\hline
		Economic Factor & Correlation Coefficient \\ [0.5ex] 
		\hline
		Unemployment Rate & 0.11\\ 
		Poverty Rate & -0.24\\ 
		Household Income & 0.17\\ 
		\hline
	\end{tabular}
	\label{Tab:Correlation}
\end{table}

We utilize the simulated joint densities to compute the systemic risk measures. 
In Table \ref{Tab:Risk-measures}, we provide the left tail systemic risk measures (CoVaR, CoES, and CoETL) on the portfolio at stress levels of 10\%, 5\%, and 1\% on the economic factors - Unemployment Rate, Poverty Rate, Household Income.
At each level, the Unemployment Rate provides the highest values for the three systemic risk measures compared to the other economic factors.
Thus, among all the stressors, the Unemployment Rate demonstrates a significantly high impact on the index.
Poverty Rate potentially has a low impact on the index according to the results of all three systemic risk measures at all stress levels.


\begin{table}[]
	\centering
	\caption{The left tail systemic risk measures (CoVaR, CoES, and CoETL) on the portfolio at stress levels of 10\%, 5\%, and 1\% on the following economic factors - Unemployment Rate, Poverty Rate, Household Income}
	\begin{tabular}{@{}lcrrr@{}}
		\toprule
		\multicolumn{1}{l}{\multirow{2}{*}{Economic Factors}} & \multirow{2}{*}{Stress Levels} & \multicolumn{3}{c}{Left Tail Risk Measures} \\ \cmidrule(l){3-5} 
		\multicolumn{1}{c}{} & & CoVaR & CoES & CoETL \\ \cmidrule(r){1-5}
		\multirow{3}{*}{Unemployment Rate} 
		&	10\%	&	-5.88	&	-8.85	&	-5.23	\\
		&	5\%		&	-9.01	&	-12.27	&	-7.06	\\
		&	1\%		&	-14.82	&	-16.32	&	-11.50	\\
		\cmidrule(r){1-5}
		\multirow{3}{*}{Poverty Rate}
		&	10\%	&	-0.67	&	-1.29	&	-0.92	\\
		&	5\%		&	-1.31	&	-2.07	&	-1.24	\\
		&	1\%		&	-2.42	&	-3.15	&	-2.05	\\
		\cmidrule(l){1-5} 
		\multirow{3}{*}{Household Income}
		&	10\%	&	-1.45	&	-2.14	&	-1.20	\\
		&	5\%		&	-2.30	&	-2.91	&	-1.66	\\
		&	1\%		&	-3.71	&	-3.86	&	-2.46	\\
		\cmidrule(l){1-5} 
	\end{tabular}
	\label{Tab:Risk-measures}
\end{table}

In conclusion, the Unemployment Rate potentially has a high impact on the financial losses due to crimes in the United States.
Hence, these findings will help investors gauge the market risk of our portfolio for hedging strategies to alleviate potential losses due to economic crashes.

\section{Discussion and Conclusion} \label{sec:DC_CI}


We proposed constructing a portfolio that outlines the financial impacts of various types of crimes in the United States.
In order to that, we modeled the financial losses of crimes reported by the Federal Bureau of Investigation
using the annual cumulative property losses due to property crimes and cybercrimes.
Then, we backtested the index using VaR models at different levels to find a proper model for implementing in evaluation processes.
As a result, we utilized ARMA(1,1)-GARCH(1,1) with Student's t model to evaluate the crime portfolio.

We presented the use of our portfolio on crimes through option pricing, risk budgeting, and stress testing.
First, we provided fair values for European call and put option prices (Figure \ref{call_price} and \ref{put_price}) and implied volatilities (Figure \ref{Implied_put}) for our portfolio.
Second, we found the risk attributed to each type of crime based on tail risk and center risk measures.
Third, we evaluated the performance of our index for the economic crisis by implementing stress testing.
According to the findings, in the United States, the Unemployment Rate potentially has a higher impact on the financial losses due to the crimes incorporated in this study compared to the Poverty Rate and Median Household Income.

The proposed portfolio on crimes is an attempt to implement a financial instrument for hedging the intrinsic risk induced by crime in the United States.
The main objective of this index is to forecast the degree of future systemic risk caused by crimes.
The findings, estimated option prices, risk budgets, and systemic risk outlined in the portfolio will help investors with financial planning on our portfolio and forewarn them to transfer insurance risk to capital market investors. 
While the portfolio on crimes is specifically constructed for the United States, it could be modified to calculate the risk in other regions or countries using a data set comparable to FBI crime data.

\clearpage
\normalem

\end{document}